# Creating tunable and coupled Rashba-type quantum dots atom-by-atom

Wouter Jolie,[1] Tzu-Chao Hung,[1] Lorena Niggli,[1] Benjamin Verlhac,[1] Nadine Hauptmann,[1] Daniel Wegner,[1] Alexander Ako Khajetoorians[1,*]

[1] Institute for Molecules and Materials, Radboud University, 6525 AJ Nijmegen, the Netherlands

*corresponding author: a.khajetoorians@science.ru.nl

**Artificial lattices created by assembling atoms on a surface with scanning tunneling microscopy present a platform to create matter with tailored electronic, magnetic and topological properties. However, such artificial lattices studies to date have focused exclusively on surfaces with weak spin-orbit coupling. Here, we created artificial and coupled quantum dots by fabricating quantum corrals from iron atoms on the prototypical Rashba surface alloy, $BiCu_2$, using low-temperature scanning tunneling microscopy. We quantified the quantum confinement of such quantum dots with various diameter and related this to the spatially dependent density of states, using scanning tunneling spectroscopy. We found that the density of states shows complex distributions beyond the typical isotropic patterns seen in radially symmetric structures on (111) noble metal surfaces. We related these to the energy-dependent interplay of the confinement potential with the hexagonal warping and multiple intra- and interband scattering vectors, which we simulated with a particle-in-a-box model that considers the Rashba-type band structure of $BiCu_2$. Based on these results, we studied the effect of coupling two quantum dots and exploited the resultant anisotropic coupling derived from the symmetry of the various scattering channels. The large anisotropy and spin-orbit coupling provided by the $BiCu_2$ platform are two key ingredients toward creation of correlated artificial lattices with non-trivial topology.**

Understanding the interplay of topology, spin-orbit-coupling and electron-mediated interactions in crystal structures and how they relate to various quantum phases of matter is one of the central goals in condensed matter physics. There are many approaches in recent years that have been developed to create designer matter, where tailored electronic or magnetic structures are created by bottom-up approaches[1-4]. Of the variety of approaches available today, scanning tunneling microscopy and spectroscopy (STM/STS) provide a powerful toolbox combining atomic-scale fabrication and on-site characterization[4, 5]. In this paradigm, impurities are patterned on a surface using atomic manipulation, and the development of tailored properties is monitored with STM/STS. For example, artificial lattices have been constructed to realize lower dimensional band structures[6-9], topological edge states[10], and synthetic Dirac quasiparticles[5, 11, 12], as well as serve as a platform for topological superconductivity[13, 14].

Quasiparticle interference (QPI) is an important aspect in creating artificial lattices based on the STM/STS approach. This is exemplified by the pioneering example of the quantum corral[15, 16], where an artificial quantum dot (QD) mimicking zero-dimensional atomic states, was created by sculpting the scattering potential stemming from impurity-induced QPI. The confinement potential of QDs made in this fashion have been utilized to sculpt the Kondo effect via the quantum mirage[17], as well as used to create coupled artificial atoms[16, 18]. More recently, this concept was expanded to create periodic lattices of coupled artificial atomic sites yielding synthetic band structures, as exemplified by the creation of synthetic graphene[5]. In this approach, patterned impurities lead to periodic scattering, where focused quasiparticles weakly interact between chosen artificial sites on the surface[19]. Nevertheless, these experiments have focused exclusively on using the quasiparticles originating from the surface state of Cu(111), which exhibits weak spin-orbit coupling[20]. The synthetic development of many classes of electronic structure requires incorporating strong spin-orbit coupling in these platforms[21]. This necessitates the development of high-quality surfaces, which exhibit large spin-orbit coupling and strong QPI, suitable for atomic manipulation.

Here, we tailored artificial Rashba-type QDs, derived from the QPI originating from electronic bands of the Rashba surface alloy $BiCu_2$ grown on Cu(111)[22, 23]. Using STM/STS, we first characterized the structural and electronic properties of individual Fe atoms. Using atomic manipulation, we then created

QDs of various sizes. We characterized the spatially dependent confined electronic states in each QD, and quantified the interplay between the confinement potential and the underlying QPI. Based on spatially dependent measurements, we observed strongly anisotropic features at particular energies, which deviate from QDs that we fabricated from CO on Cu(111). Using a particle-in-a-box model and comparing to spatially dependent spectroscopy, we traced the rich behavior of the confined wave functions to the interplay between multi-band scattering, hexagonal warping, and the size of the QD. Using the hexagonal symmetry of the $BiCu_2$ system, we constructed coupled QD pairs that take advantage of anisotropic coupling. Our experiments introduce an electronic platform with large anisotropy and spin-orbit coupling, two key ingredients to create correlated artificial lattices with non-trivial topology.

Upon deposition of Bi on Cu(111) and annealing (cf. SI for further experimental details), Bi atoms embed in the topmost Cu(111) layer and form a well-ordered ($\sqrt{3}\times\sqrt{3}$)R30° superstructure ($BiCu_2$)[24]. The surface alloy can be imaged using STM constant-current imaging (Fig. 1(a)). $BiCu_2$ features two strongly confined, hole-like surface states, which exhibit a giant Rashba effect due to large spin-orbit coupling[25-27]. While the inner band is almost isotropic, the outer band shows pronounced hexagonal anisotropy, leading to increased nesting along specific directions in momentum space. This can be directly imaged using QPI as well as with angular resolved photoemission[22, 23, 25-27]. A sketch of the band structure is depicted in Fig. 1(d-e), including the possible spin-conserving QPI scattering vectors.

In the subsequent experiments, we created large terraces of high-quality $BiCu_2$, followed by deposition of individual Fe atoms on the surface to create well-defined point scatterers. Fe atoms appear in constant-current imaging (Fig. 1(a)) as an almost round protrusion. By imaging the surrounding lattice, we found that Fe preferably adsorbs on top of Cu atoms, as depicted in the corresponding structural model in Fig. 1(b). Fe is surrounded by three Bi atoms, leading to two equivalent adsorption sites with three-fold symmetry and to an apparent triangular deformation from the round shape of the Fe atom (Fig. 1(a)). We measured point spectra (d$I$/d$V$) of individual Fe atoms, in comparison to the $BiCu_2$ substrate, as illustrated in Fig. 1(c). Spectra of bare $BiCu_2$ exhibit a pronounced peak caused by a van Hove singularity at $V_S$ = 0.25 V, resulting from the onset of the inner Rashba-split $sp_z$ band[23, 28]. In comparison, individual Fe atoms are rather featureless in the probed energy window, aside from a pronounced Fano resonance at the Fermi energy. While this feature is not at the focus of this manuscript, we attribute it to a Kondo resonance[15, 29], with a half-width at half maximum (HWHM) of roughly $\Gamma = 10$ mV. Therefore, we consider the Fe atoms as hard-wall scatterers, and neglect the contributions to the QPI induced by the Kondo-like feature. As a result, we normalized all spectra taken in the generated QDs by dividing them with the substrate reference spectrum.

In Fig. 2, we detail constant-current imaging of the local density of states (LDOS) at various energies of two circular QDs with different radii ($R$ = 7.3 nm (a) and 6.15 nm (b)). As Fe atoms exhibit strong QPI, the formation of the QD leads to a strong quantum confinement potential. The patterns imaged at a given energy inside the QD are related to the interference of standing waves originating from the various scattering channels ($q_i$) in the band structure (Fig. 1(d-e)). Reducing the radius of the QD pushes the energy of the confined states further away from the onset energy of the surface state, as expected from a particle-in-a-box model. In the original quantum corral work by Crommie et al.[15], this led to circular fringes at all energies, stemming from one isotropic scattering channel, related to the dispersion of the Shockley surface state of Cu(111). In comparison here, more complex patterns were observed within a given QD, depending on $V_S$. For $R$ = 7.3 nm (Fig. 2(a)), we found circular-like fringes near the onset of the inner band. Going lower in energy (i.e., following the downward dispersion), there was a strong deviation from circular fringes, exemplified by the emergence of hexagonal patterns within the circular confinement potential featuring a non-trivial spatial distribution within the QD. Intriguingly, we also found that some of the multiple hexagonal patterns at a given energy are rotated by 30° with respect to each other (e.g. $V_S$ = 125 mV). In comparison, the QD with radius $R$ = 6.15 nm exhibited noticeably different distributions, compared to the previous QD. The emergence of hexagon patterns can be seen closer to the band onset, whereas the interplay of different hexagonal fringes with different apparent symmetry, compared to the $R$ = 7.3 nm, leads to vastly different wave functions at energies further away from the band onset.

In order to understand the origins of these patterns, we first revisit the possible scattering vectors stemming from the BiCu$_2$ band structure[23]. At a given energy, the QPI patterns are dominated by a set of scattering wave vectors that connect two states in the band structure (Fig. 1(d)). Here, we assume the outer $p_xp_y$ band (with large wave vectors) to be spin-degenerate close to the Fermi energy, in line with angular-resolved photoemission measurements[22, 25]. In contrast, the inner band is strongly spin-split due to the Rashba effect. This leads to four possible spin-conserving scattering wave vectors: two intraband scattering wave vectors $q_1$ and $q_4$, which connect states within the same band, and two interband scattering wave vectors $q_2$ and $q_3$ that connect states between inner and outer bands, respectively. The $k$-dependence of these scattering wave vectors at a fixed energy is shown in Fig. 1(e), which displays a constant-energy contour of the band structure in momentum space. While $q_1$ is isotropic, $q_2$ - $q_4$ inherit the hexagonal warping of the outer band.

We subsequently measured d$I$/d$V$ point spectra along the diameter of a QD and compared them to calculations based on a particle-in-a-box model (Fig. 3). Each spectrum was taken after moving to a subsequent point along the line, with the feedback loop enabled, and then measuring the point spectrum with the feedback loop opened at the stabilization voltage. Fig. 3(b) shows the resulting normalized d$I$/d$V$ intensity in a false-color plot as a function of $V_S$ and position along the diameter. The regions of high intensity can be associated with the LDOS of the localized wave functions ($|\psi|^2$), which are related to the various states of the particle in the box. The data reveals a discrete set of standing waves inside the QD, starting with the first eigenstate close to the band maximum of the inner Rashba-split band, and with the wavelength of the higher states decreasing with decreasing energy, as expected for hole-like bands.

To simulate the particle in a box, we simplified the confinement potential generated by individual Fe atoms using an infinite potential. In order to account for the finite size of each Fe atom, we considered a circular potential of the QD with a theoretical radius reduced by the atomic radius of Fe (0.15 nm). The solutions of the Schrödinger equation are $l^{th}$ order Bessel functions, $J_l$, defined by a set of quantum numbers $(n, l)$ with momentum $k_{n,l} = z_{n,l}/R$ and energy $E = \frac{\hbar^2 k_{n,l}^2}{2m^*}$, with $m^*$ the effective mass of the nearly free surface band[15]. The resultant LDOS is given by $J_l^2/N_{n,l}$, where $N_{n,l}$ ensures proper normalization. To verify our model, we also measured the situation for a QD built from CO molecules on Cu(111), and simulated it using a single isotropic scattering channel based on the Cu(111) surface state (Supplementary Figure 1). In this case, we found near reasonable agreement between the experimental results and the model. For BiCu$_2$, the presence of multiple scattering wave vectors necessitates considering more channels, each with a given contribution $w(q_i)$ to the LDOS (see below). We considered the four scattering wave vectors $q_1$ - $q_4$ to account for all the features observed in the experiments, and treated each of them in an independent particle-in-a-box problem, respectively (Supplementary Figure 2). To be able to simulate these wave vectors using a free-particle Hamiltonian, we analytically described the dispersions of all four scattering wave vectors using parabolic fits of the underlying bands to obtain the onset energies $E_{0,q}$ and effective masses $m_q^*$, (see supporting information). This results in a set of wave functions $(n,l)_q$ for all four confined scattering wave vectors. Based on previous QPI measurements from native defects[23], the relative strength of each scattering channel may vary. We therefore weighed the contribution of all four channels, using experimental input, such that $w(q_1) = 5w(q_2) = 5w(q_3)$ (Supplementary Figure 3). We note that $q_3$ contributions have not been observed before in QPI of native defects[23], but were relevant to describe these experiments. This most likely stems from differences in the scattering potential between native defects and individual Fe atoms. Finally, we neglected any energy dependence of these weights in this simulation. Using the simulations as a basis, we could ultimately identify the quantization conditions for a given QD. The dominating $(n,l)$ eigenstate contributions of the inner band ($q_1$) are indicated in Fig. 2.

The resulting simulation for the QD with $R = 7.3$ nm is plotted in Fig. 3(c). We find that most of the intensity in the unoccupied states stems from the isotropic inner band ($q_1$). Nevertheless, there are non-trivial contributions from $q_2$ and $q_3$, leading to the deviation from circular fringes. For instance, the $x$-dependent energy and broadening of the state near $V_S = 50$ mV in the center of the QD is found both experimentally and theoretically and can be traced back to a local superposition of $q_1$ and $q_3$ states. Further fingerprints of $q_3$ are discussed in the supporting information. Similarly, a wealth of additional whispering gallery modes is found in the simulation near the rim[30]. We note that the addition of $q_4$ contributions did not lead to further improvement of the agreement between experiment and model (Supplementary Figure 4). While the inclusion of interband scattering ($q_2$ and $q_3$) helps to improve the comparability between simulation and experiment for the unoccupied states, there are more qualitative differences in the occupied region. For instance, the alternation between nodes and antinodes in the center of the QD from one confinement state to the next, as expected from a particle-in-a-box model, is not followed when crossing the Fermi level: the confined states seen near $V_S = 50$ mV and $V_S = -25$ mV both have antinodes. The variation may result from the assumption of a constant infinite scattering potential, which may be too simplified. Recent QPI measurement on Re(0001) revealed a sign change of the scattering potential when crossing $E_F$, which was attributed to a different scattering behavior of holes and electrons[31]. In contrast, measurements of the Ag(111) surface state showed no sign change at $E_F$[32], in agreement with our measurements of QDs on bare Cu(111) (see supporting information). In addition to a change in sign, hybridization of the Fe atomic states with the substrate may also create an energy-dependent scattering potential, which may effectively absorb and/or transmit part of the scattered waves. We note that at larger negative energy, deeper into the occupied bands, we found a tendency toward more uniform hexagonal type patterns in the QD (Supplementary Figure 5). We also considered QDs with different radii, both experimentally and theoretically (Supplementary Figure 6).

As shown in numerous studies, QPI can be utilized to create artificial lattices[5, 10, 12, 19, 33-35]. Within this paradigm, a periodic array of scatterers is used to create an anti-lattice, which focuses quasiparticles at periodic positions. These focal points define an artificial lattice site. In this way, a QD can be identified as one artificial atom[18]. All previous experiments based on the QPI concept used quasiparticles originating from the Cu(111) surface state, which exhibits weak spin-orbit coupling. There is a strong desire to introduce spin-orbit coupling, like Rasha-type coupling, into these artificial systems to create topological and correlated matter[21]. Nevertheless, this strategy depends on isotropic scattering stemming from a single scattering channel, and neglects additional effects such as bulk scattering or energy-dependent potential scattering from the impurities. Using $BiCu_2$ as a prototypical Rashba platform, the design of artificial lattices is complicated by the compilation of effects demonstrated above: anisotropic scattering, a complex scattering potential, and the presence of multiple scattering channels. For example, the presence of multi-band scattering will ultimately delocalize artificial lattice sites, due to the different modulations present, as seen in Fig. 2.

In order to understand how to couple artificial atoms in this vain, we created QD pair structures connected via a weak link in the structure to induce a wave function overlap[18]. The building block is a QD with $R$ = 3.1 nm built from 24 Fe atoms (Fig. 4(a)). We removed three Fe atoms at one side of the circular QD and fused a second constructed QD of same size next to it. This way, we constructed two nearly equivalent QD pairs, which are rotated by 30° with respect to each other (Fig. 4(b,c)), and we compare their respective electronic properties to each other as well as to the isolated QD.

Starting with the single QD, we measured STS spectra along a line as done for the previous QDs (Fig. 4(a)). The corresponding position- and energy-dependent normalized d$I$/d$V$ intensity is shown in Fig. 4(d). Similar to the larger QDs in Fig. 3, we found a set of standing waves, but now the energy spacing between resonances is larger due to the smaller radius. The corresponding model simulation is shown in the supporting information (Supplementary Figure 6). We compare these spectra with the results obtained across the QDs coupled along the BiCu$_2$ [100] direction (Fig. 4(e)). While we can identify the resonances found in the isolated QD at roughly the same energies, small energy shifts are evident in the coupled case, breaking the mirror symmetry found in the isolated QD. The resonances are moving up in energy close to the junction between the QDs, while moving down in energy close to the outer rims. The same trend is also found for the coupled QD oriented along [110].

We confirm the effects of coupling in these QD pairs by mapping the LDOS at the energies marked with dashed lines in Fig. 4(d-e). The resulting constant-current d$I$/d$V$ maps are shown in Fig. 5. Using these maps, we can clearly identify changes to the localized wave functions, in comparison to the single QD case. For example, at $V_S$ = 200 mV (Fig. 5(a)), the single QD map features an almost circular ring close to the center, and minimal LDOS close to the rim. In both QD pairs, the ring becomes severely deformed, redistributing LDOS toward to the junction region. In addition, a strip of LDOS connects the two rings through the junction, indicative of a coupling between to two QDs. In contrast, maps taken at $V_S$ = 166 mV (Fig. 5(b)) feature a broad LDOS in the QD center, which is unaffected in the QD pair structures. Maps at energies where small wavelengths dominate the LDOS reveal signs of bonding and antibonding character. At $V_S$ = -275 mV (Fig. 5(c)) the QD pair coupled along the [100] direction showed maximum intensity in the center of the junction, while at $V_S$ = -375 mV (Fig. 5(d)) we observed a local minimum. These features are reminiscent of the wave functions expected from bonding and antibonding states, with antinodes and nodes in the bond center, respectively.

Finally, we address the orientation dependence of the coupling between the two QDs. As discussed above, the superposition of isotropic and hexagonal standing waves in BiCu$_2$ leads to anisotropic patterns in isolated QDs. We found strong evidence for a resulting anisotropic coupling when comparing

d$I$/d$V$ maps of QD pairs coupled along [100] vs. [110] (Fig. 5(c,d)). The standing wave patterns are qualitatively different in the junction region. The QD pair coupled along [110] shows an antinode in the junction at $V_S$ = -275 mV (Fig. 5(c)) and a node at $V_S$ = -375 mV (Fig. 5(d)), i.e., the situation is reversed compared to the QD pair coupled along [100].

In conclusion, we constructed Rashba-type artificial quantum dots using atomic manipulation of Fe atoms to form circular confinement potentials on the BiCu$_2$/Cu(111) surface alloy. Scanning tunneling spectroscopy revealed anisotropic multi-band QPI, i.e., the spatial distribution of the wave functions was found to be much more complex, when compared to quantum dots fabricated on surfaces with isotropic single-band scattering. We investigated the effects of hexagonal warping and multi-band scattering as a function of energy for QDs with various radii. The observed complex spatial distributions of the wave functions could be partially captured by a nearly-free particle in a box model considering the underlying band structure. There are clear discrepancies, particularly at higher energies in the occupied states, which requires a more detailed theoretical treatment, for example by considering the energy-dependent scattering potential of the Fe atoms, the spin-orbit coupling, and the energy-dependent variations in the quasiparticle scattering. Due to the complexity of the scattering in this system, and therefore the non-trivial nature of how to design an anti-lattice for artificial lattices, we considered how to couple QDs using this platform by constructing dimers of quantum corrals. We demonstrated that coupled states can be identified in QD pair structures, revealing anisotropic coupling. These results serve as the basis for understanding how to incorporate Rashba-type coupling for artificial lattices constructed on surfaces, including the role of multi-band scattering and hexagonal warping.

**Acknowledgements**

We would like to thank Cristiane Morais Smith and Sander Kempkes for extended theoretical discussions. We also thank Raffael Spachtholz for experimental support. This project has received funding from the European Research Council (ERC) under the European Union's Horizon 2020 research and innovation programme (grant agreement No 818399). We acknowledge NWO-VIDI project "Manipulating the interplay between superconductivity and chiral magnetism at the single-atom level" with project number 680-47-534. WJ acknowledges support from the Alexander von Humboldt

Foundation via the Feodor Lynen Research Fellowship. B.V. acknowledges funding from the Radboud Excellence fellowship from Radboud University in Nijmegen, the Netherlands.

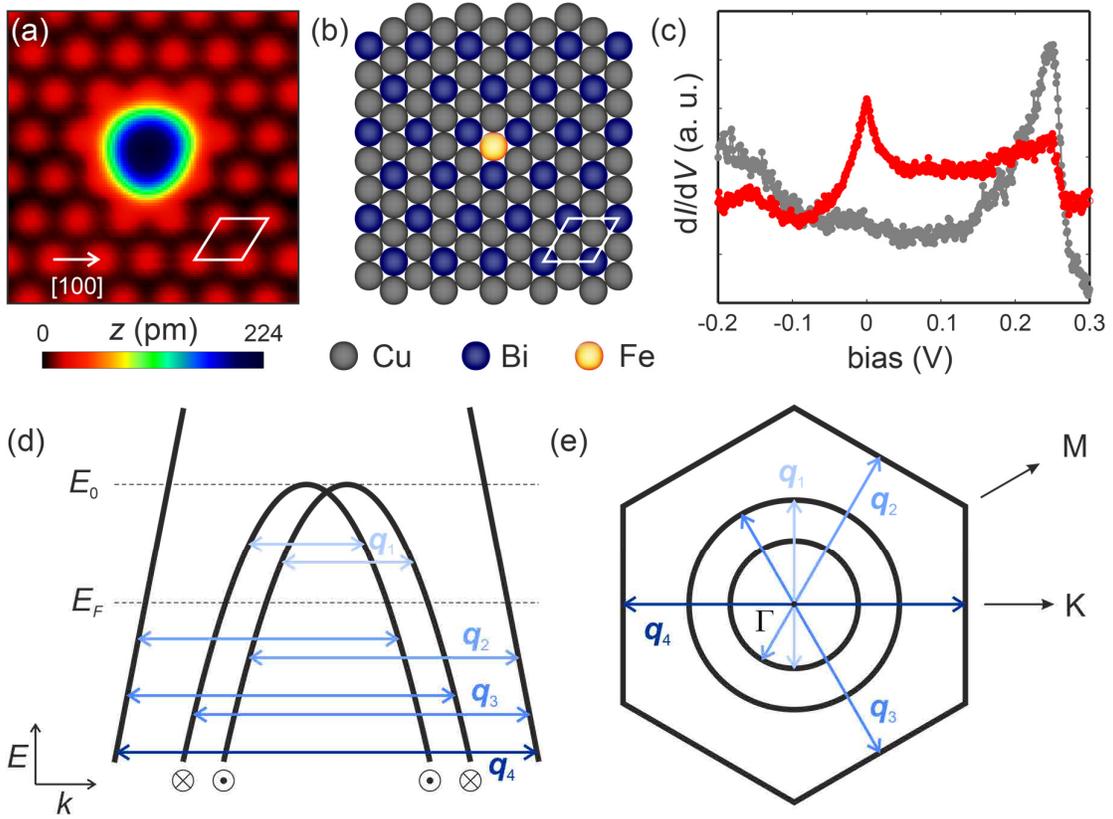

**Figure 1. Single Fe atom on BiCu$_2$**: (a) Constant-current STM image of a single Fe atom on BiCu$_2$ ($V_S$ = 5 mV, $I_t$ = 100 pA). (b) Structural model of (a), revealing the adsorption site of Fe. (c) Point spectra taken on BiCu$_2$ (gray) and Fe (red), respectively. (d) Sketch of the hole-like surface states of BiCu$_2$. The Fermi energy $E_F$ and onset $E_0$ of the inner Rashba-split band are denoted. Blue colored arrows represent intraband ($q_1$ and $q_4$) and interband ($q_2$ and $q_3$) scattering vectors, which lead to visible quasiparticle interference. (e) Simplified cartoon of the constant-energy contour of BiCu$_2$ in momentum space. The inner Rashba-split band shows circular symmetry, while the outer band shows hexagonal symmetry.

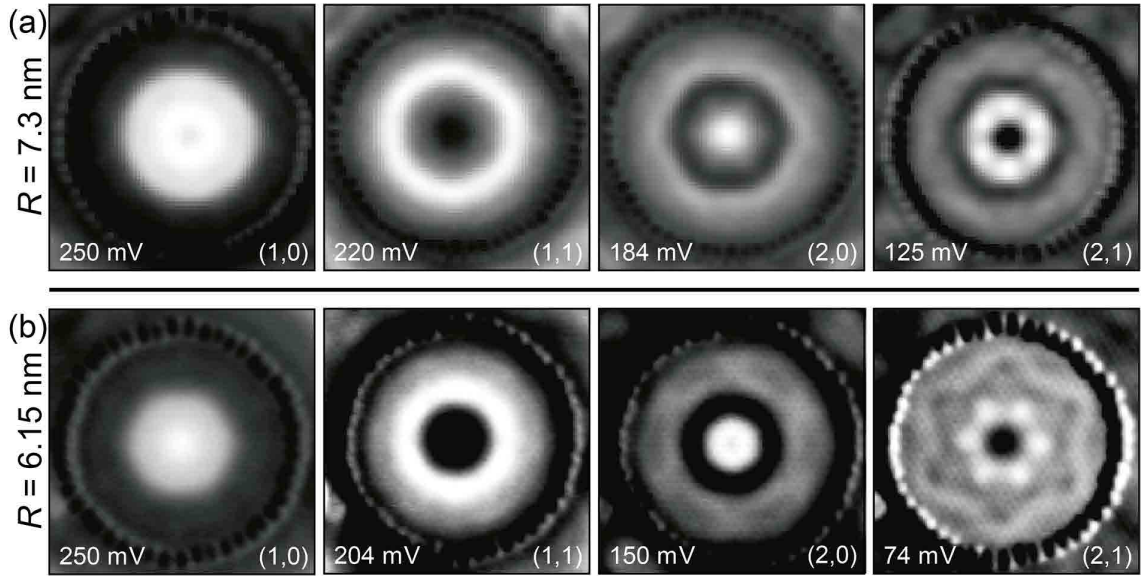

**Figure 2. Confined states of the Rashba QD with tunable anisotropy for two different radii.** (a) Real-space constant-current d$I$/d$V$ maps of a QD with $R$ = 7.3 nm built from 56 Fe atoms. Labels in the lower right corner refer to the dominating (*n,l*) eigenstate contribution of the inner band ($q_1$). The superposition with eigenstates of the outer band as well as interband eigenstates creates patterns with clear hexagonal symmetry ($q_2$ - $q_4$). While the (2,0) state points in the [100] direction with respect to the Bi superstructure (horizontal), the inner ring of the (2,1) state points along [110] (vertical). (b) Real-space constant-current d$I$/d$V$ maps of the QD with $R$ = 6.15 nm built from 48 Fe atoms. The symmetry of the (2,0) and (2,1) states has changed. While the (2,0) state points along [110], the inner ring of the (2,1) state points along [100]. (Standard linear grayscale mapping was used for all images, and adjusted individually to maximize contrast).

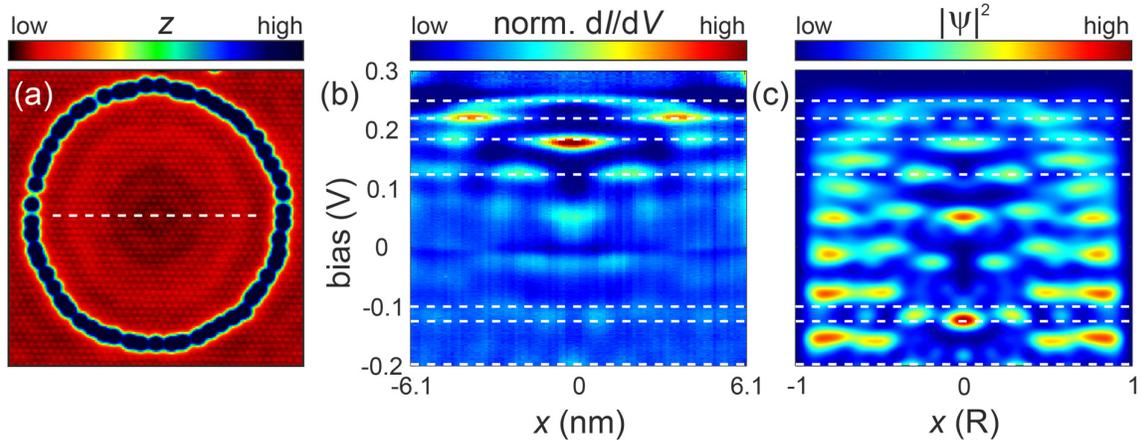

**Figure 3. Dispersion of an artificial Rashba QD**: (a) Constant-current STM image of a QD built from 56 Fe atoms with $R = 7.3$ nm ($V_S = 5$ mV, $I_t = 100$ pA). (b) Set of normalized STS spectra measured along the cross-sectional line indicated in (a). (c) Particle-in-a-box simulation of the QD, containing contributions from $q_1$, $q_2$ and $q_3$ with fixed weights (w($q_1$) = 5 w($q_2$) = 5 w($q_3$)). The horizontal dashed lines denote the energies of the maps shown in Fig. 2(a) and Fig. S4.

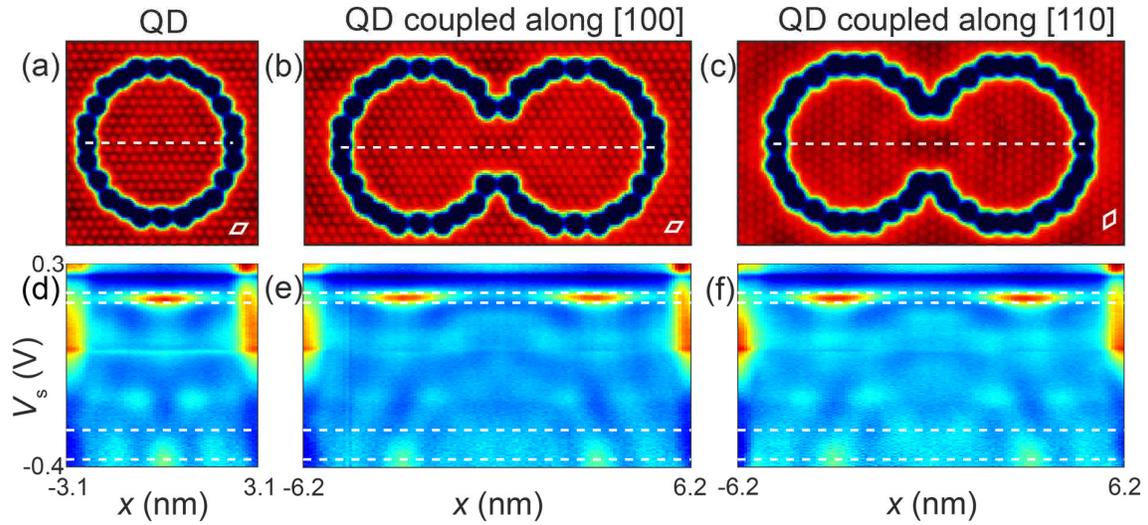

**Figure 4. Anisotropic coupling of quantum dot pairs:** (a) STM constant-current image of a small ($R$ = 3.1 nm) QD ($V_S$ = 5 mV, $I_t$ = 20 pA). (b) STM image of a QD pair coupled along [100] ($V_S$ = 5 mV, $I_t$ = 20 pA). (c) STM image of a QD pair coupled along [110] ($V_S$ = 10 mV, $I_t$ = 100 pA). (d-f) Set of normalized spectra measured along the line denoted in (a)-(c), respectively.

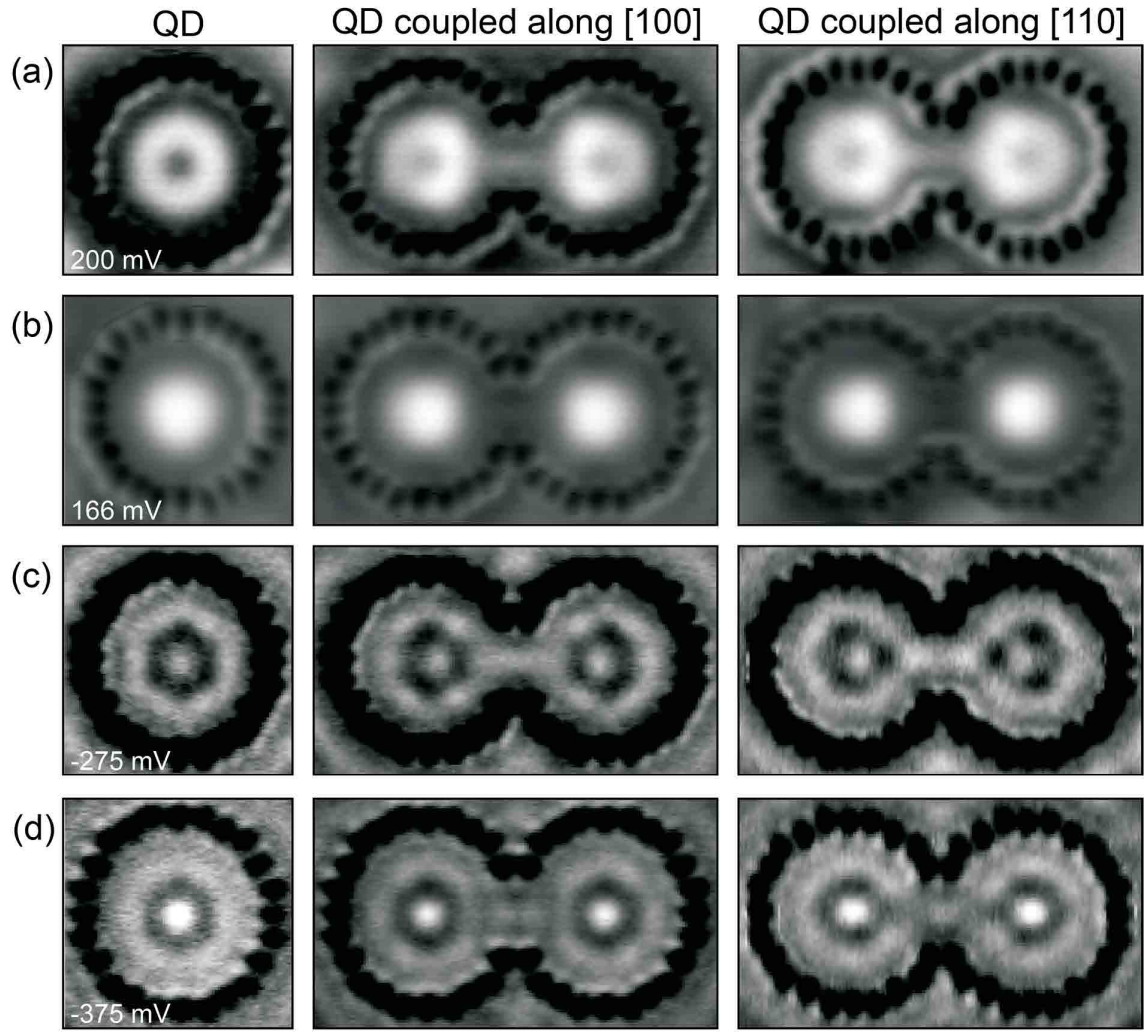

**Figure 5. Anisotropic coupling in artificial QD pairs:** (a) Maps taken at $V_S$ = 200 mV of the three structures shown in Fig. 4. LDOS intensity shifts towards the junction, forming a bonding state. (b) Maps taken at 166 mV indicate no coupling at this energy. (c) Maps taken at -275 mV. The QD pair coupled along [100] exhibits a maximum in the LDOS intensity at the junction center, while that coupled along [110] has a minimum. (d) Maps taken at -375 mV. Now the QD pair coupled along [100] exhibits a minimum in the LDOS intensity at the junction center, while the QD pair coupled along [110] has a maximum.

# Supporting information: Creating tunable and coupled Rashba-type quantum dots atom-by-atom

Wouter Jolie,[1] Tzu-Chao Hung,[1] Lorena Niggli,[1] Benjamin Verlhac,[1] Nadine Hauptmann,[1] Daniel Wegner,[1] Alexander Ako Khajetoorians[1]

[1] *Institute for Molecules and Materials, Radboud University, 6525 AJ Nijmegen, the Netherlands*

## 1. Methods

The experiments were performed in ultra-high vacuum (UHV) using a commercial UHV LT-SPM system (Createc). A Cu(111) single crystal was repeatedly cleaned with ion bombardment (Neon, 1.5 keV) and subsequently annealed to 450°C for 10 minutes. Bi was evaporated on the Cu(111) surface held at 100°C, followed by an annealing step of 10 minutes at 150°C. This procedure led to large, clean terraces of the $BiCu_2$ surface alloy. Single Fe atoms were deposited on the surface held at 8 K inside the cryogenic STM. All STM/STS measurements were performed at 5 K. Atomic manipulation of Fe atoms was performed with a tunneling current of ~30 nA and a bias voltage of 5 mV. For scanning tunneling spectroscopy (STS), we use a lock-in technique with a modulation of 1-2 mV for point spectra and 2-10 mV for d$I$/d$V$ maps, with a frequency of 433-777 Hz.

## 2. Quantum dots made of CO on Cu(111)

To test our particle-in-a-box model, we built two quantum dots (QDs) on the Cu(111) surface using CO molecules, similar to Ref. [1], and two corrals built from Fe atoms on Cu(111)[2, 3]. The results are shown in Fig. S1. The corresponding experimentally recorded confined states are shown in Fig. S1(b,e). As expected for a particle-in-a-box model for an isotropic quantum corral, we found an increase of the number of nodes and antinodes, as well as an alternation between nodes and antinodes in the center of the QD from one confinement state to the next, with no change when crossing $E_F$. For the simulation, we used a single, electron-like band which is manually fit to the experimental *dI/dV* signal. The theoretical radius was reduced by 0.25 nm to account for the finite size of the CO molecules[4]. In line with the experiments, we found an alternating sequence of maxima and minima in the center of the corral. Overall, the agreement between experimental and simulated results is strong in all energy ranges probed, in contrast to what was observed in the main manuscript for the corrals built on BiCu$_2$. For the smaller QD, the most obvious discrepancy is that the eigenstates are broader. We attribute this to the finite scattering potential of the CO molecules.

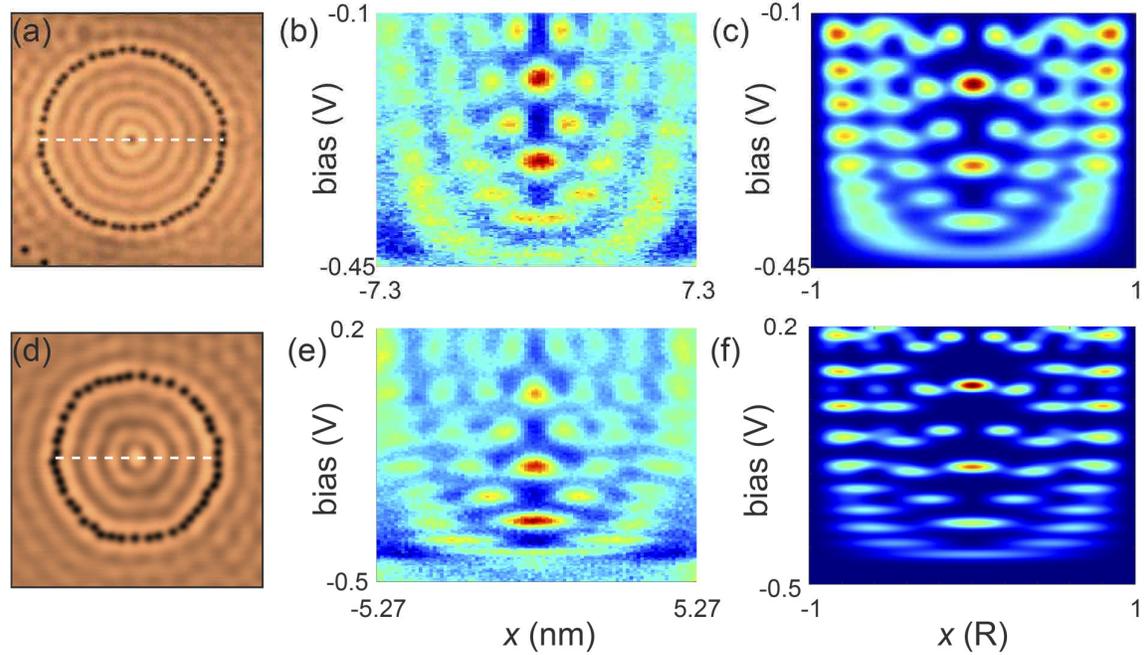

**Figure S1.** (a) Constant-current STM image of a QD with radius $R$ = 7.3 nm ($V_S$ = 50 mV, $I_t$ = 200 pA), formed from CO molecules on Cu(111). (b) Set of normalized STS spectra measured along the cross-sectional line indicated in (a). (c) Simulation using $E_0$ = -0.44 eV and $m^*$ = 0.44 $m_e$. (d) Constant-current STM image of a quantum corral with radius $R$ = 5.27 nm ($V_S$ = 50 mV, $I_t$ = 200 pA) formed from CO molecules on Cu(111). (e) Set of normalized STS spectra measured along the cross-sectional line indicated in (d). (f) Simulation using $E_0$ = -0.44 eV and $m^*$ = 0.44 $m_e$.

### 3. Approximated band structure of BiCu$_2$

Standing waves in BiCu$_2$ result from four possible scattering vectors ($q_i$). We fit the dispersion of these scattering wave vectors using parabolic fits and obtained an approximated effective band structure of BiCu$_2$, shown in Fig. S2. These effective bands are used to simulate the confined states of BiCu$_2$. The red band stems from the inner $sp_z$-type Rashba-split surface state, the green and dark blue bands simulate interband scattering wave vectors. The light blue band stems from the outer $p_xp_y$-band. We find single parabolic dispersions for both the inner and the outer bands. For the inner band, there are two degenerate dispersions present for this contribution, i.e., the Rashba splitting is not visible via intraband scattering[5]. The Rashba effect, however, is still incorporated in our simulations and leads to two inequivalent interband scattering-related parabolic dispersions, which would collapse to a single dispersion without Rashba spin-orbit coupling. The outer band is treated as spin-degenerate, in accordance with ARPES and DFT results[6-9].

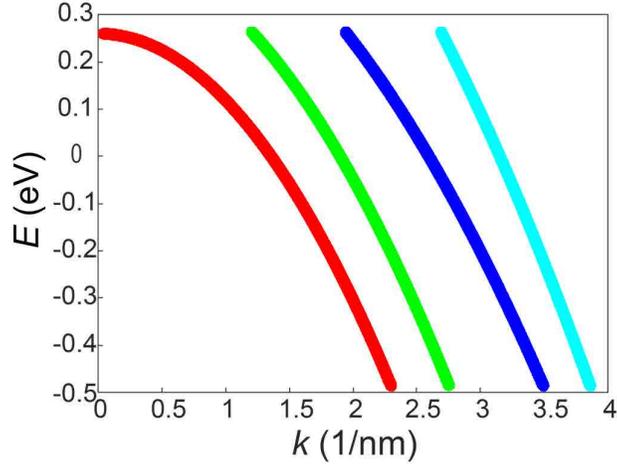

**Figure S2.** Effective parabolic bands used to simulate the possible scattering wave vectors within the band structure of BiCu$_2$. The parameters used for all QD simulations on BiCu$_2$ are $E_1$ = 0.26 eV, $m_1$ = -0.27 $m_e$ (red); $E_2$ = 0.441 eV, $m_2$ = -0.312 $m_e$ (green); $E_3$ = 0.598 eV, $m_3$ = -0.429 $m_e$ (dark blue); $E_4$ = 0.971 eV, $m_4$ = -0.39 $m_e$ (light blue).

With the parabolic dispersions, we treated each band independently using a particle-in-a-box model. The solutions are Bessel functions, which are characterized by two quantum numbers, ($n$,$l$): ($n$,0) and ($n$,1) resembles the one-dimensional particle-in-a-box solutions, with maxima and minima in the center of the QD for the ($n$,0) and the ($n$,1) states, respectively. For increasing $l$, the wave function becomes more localized toward the rim of the QD. We included all states present in the corresponding energy window. An additional (Gaussian) energy broadening is included, to simulate various effects of broadening (FWHM = 26 mV). All wave functions are normalized independently to 1. To be able to weigh the contribution of each individual band, we multiplied the square of each wave function for each corresponding band with a weighting factor $w(q_i)$. The superposition of all wave functions is what is plotted in the simulated plots.

## 4. Disentangling the scattering wave vector contributions

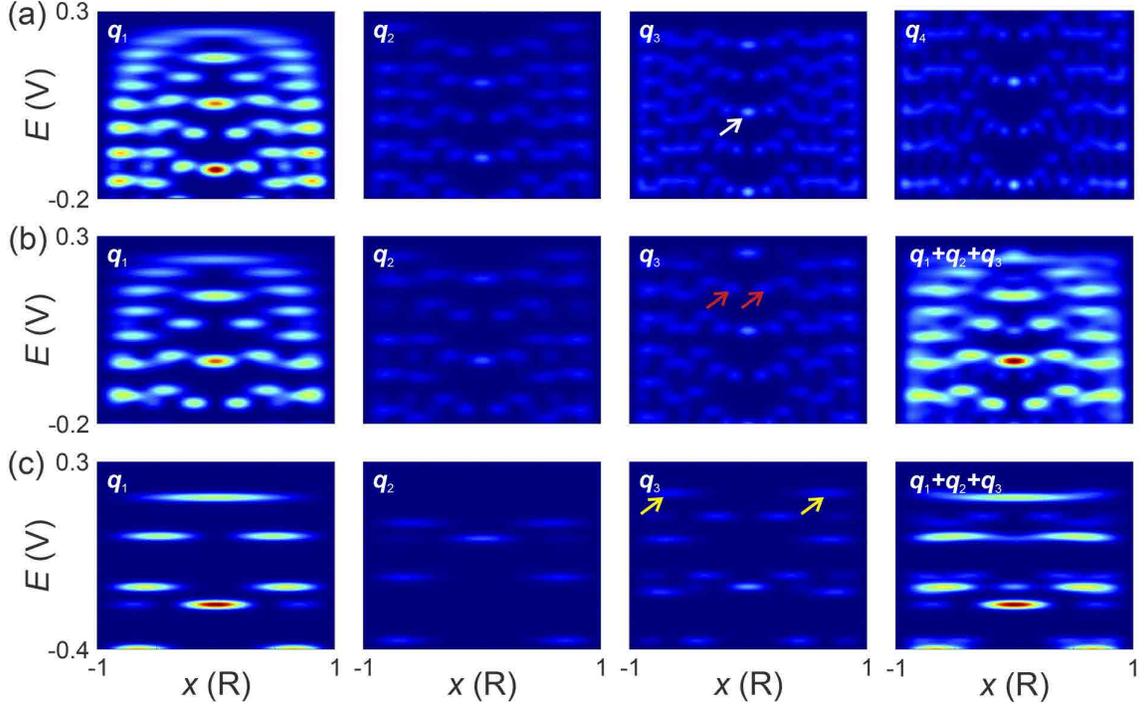

**Figure S3.** Disentangling the contributions to the local density of states: (a) Simulations of all effective bands present in the QD with $R = 7.3$ nm. The weights are $w(q_1) = 5$ $w(q_2) = 5$ $w(q_3) = 5$ $w(q_4)$. (b) Simulations of the QD with $R = 6.15$ nm. The weights are $w(q_1) = 5$ $w(q_2) = 5$ $w(q_3)$; $w(q_4) = 0$. (c) Simulations of the QD with $R = 3.1$ nm. The weights are $w(q_1) = 5$ $w(q_2) = 5$ $w(q_3)$; $w(q_4) = 0$.

To be able to attribute features present in the experimental and theoretical spectrum to a certain confined state, we plot in Fig. S3(a) the contributions of all wave vectors in the large QD ($R = 7.3$ nm), respectively. We use the same weights for $q_1$- $q_3$ as in the main manuscript. A potential contribution of $q_4$ is included as well for the sake of completeness. The broadening in the center around 50 mV can only be explained with the confined states of $q_3$, since it has a maximum at this energy, in contrast to $q_2$ and $q_4$. The location is highlighted with a white arrow. Fingerprints of the confined states of $q_3$ are found in the smaller quantum corrals as well. Fig. S3(b) shows the contributions corresponding to the QD with $R = 6.15$ nm. The two red arrows point to the state that is responsible for the ring-like shape near the center of the map at 150 mV in Fig. 2(b). From the contribution of $q_1$ alone, one would expect a maximum in the center. For the quantum corral with $R = 3.1$ nm, we found a ring-like state at 200 mV (see Fig. 5(a)), which is also found in the simulations of $q_3$, see the yellow arrows in Fig. S3(c). Overall, we found multiple instances where details in the observed results can only be understood by inclusion of $q_3$. This is very different from our test-bed QDs on Cu(111).

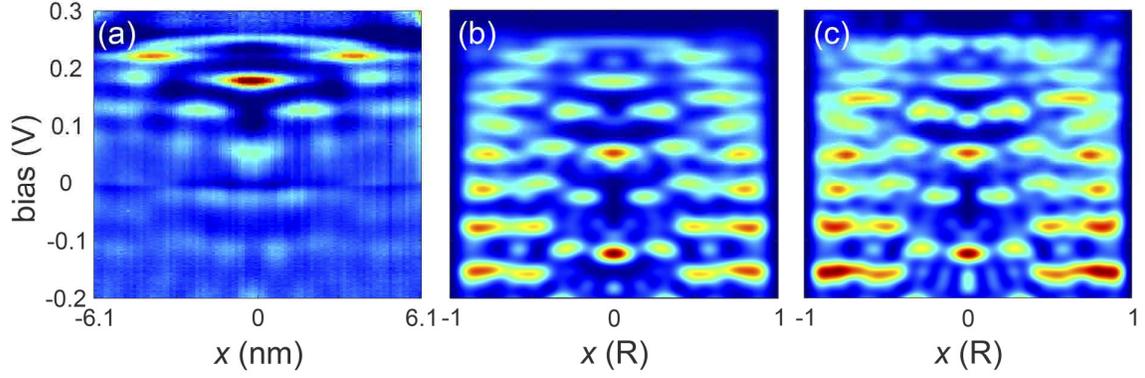

**Figure S4:** Including $q_4$ to the simulation: (a) Set of normalized STS spectra presented in Fig. 3 (b) of the main manuscript. (b) Particle-in-a-box simulation of the QD, containing contributions from $q_1$, $q_2$ and $q_3$ with fixed weights ($w(q_1) = 5\ w(q_2) = 5\ w(q_3)$). (c) Simulation containing contributions from $q_1$, $q_2$, $q_3$ and $q_4$ with fixed weights ($w(q_1) = 5\ w(q_2) = 5\ w(q_3) = 5\ w(q_4)$).

Figure S4 compares the STS spectra measured across the QD with $R = 7.3$ nm with two simulations without (Fig. S4b) and with (Fig. S4c) the contribution of $q_4$, respectively. Additional modulations are found in the simulation that includes $q_4$. For example, the confined state close to the band onset (near 250mV) shows a pronounced beating. An extra maximum is found close to 110 mV in the center of the QD. These features were, however, not found in the experiment. In addition, we did not see standing waves above the onset of the inner $sp_z$-type Rashba-split surface state in the experiments, where only features related to $q_4$ could contribute. From these findings, we conclude that the impact of $q_4$ is negligible, which is why we excluded all contributions from $q_4$.

## 5. Additional maps below the Fermi energy

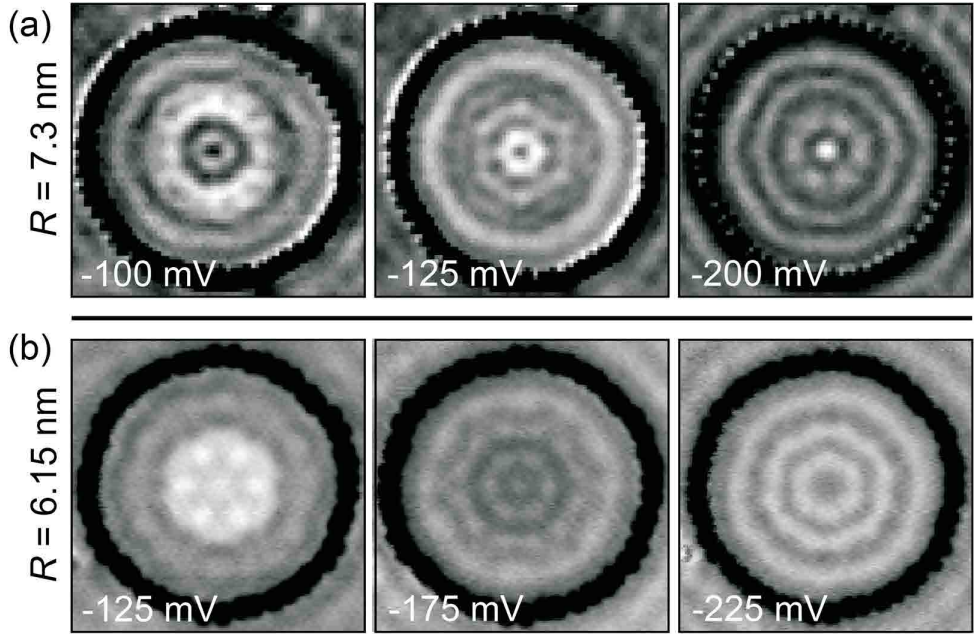

**Figure S5.** Confined states below the Fermi energy: (a) Constant-current d$I$/d$V$ maps measured at negative bias for the QD with $R$ = 7.3 nm. (b) Constant-current d$I$/d$V$ maps measured at negative bias for the quantum corral with $R$ = 6.15 nm.

Fig. S5 shows additional standing wave patterns of the two QDs presented in Fig. 2, in the occupied region, where our model does not reproduce the experimental spectra. In Fig. S5(a), again an energy-dependent rotation can be seen when comparing the symmetry of the maps at -125 mV and -200 mV. In comparison, this symmetry changed for the smaller QD shown in Fig. S5(b).

## 6. Additional electronic characterization of the smaller QDs

Fig. S6 presents complementary energy- and position-dependent LDOS data of the QDs analyzed in the main text. We start with the smallest QD we investigated in Fig. 4 of the main text, see Fig. S6(a). We show here the cross section along [110] (Fig. S6(b)), complimentary to that shown along [100] in Fig. 4(d). There was no change in the energies of the QD eigenstates when comparing both directions, as expected for an isolated quantum corral. The hexagonal anisotropies that were evident in the d$I$/d$V$ maps (see Fig. 5) can also be seen in slight differences of the spatial distribution of the eigenstates. The corresponding simulation for this QD is shown in Fig. S6(c) for the sake of completeness. Again, the comparability with the experiment is better above $E_F$, and unsatisfactory below $E_F$. An STM image of the quantum corral with $R = 6.15$ nm (see Fig. 2(b) in the main text for corresponding d$I$/d$V$ maps) is presented in Fig. S6(d). The QD states found therein are plotted in Fig. S6(e), together with the corresponding simulation in Fig. S6(f).

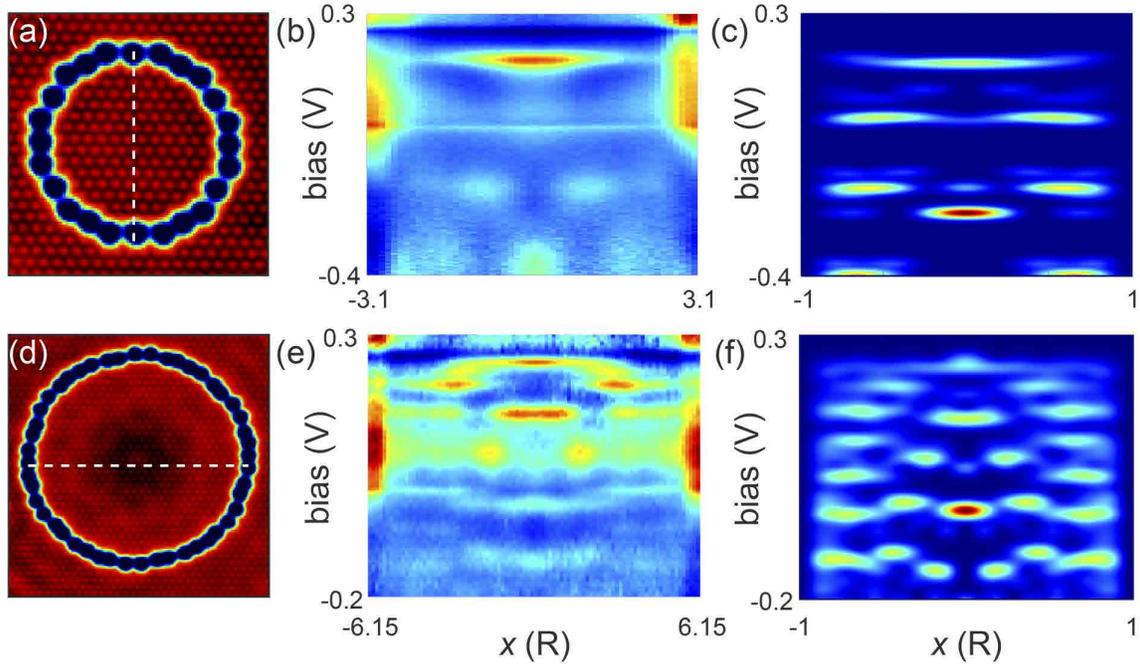

**Figure S6.** (a) STM image of an artificial QD with $R = 3.1$ nm ($V_S = 5$ mV, $I_t = 20$ pA). (b) Set of normalized d$I$/d$V$ spectra measured along the line denoted in (a), which is the [110] direction. A set of confined states appears inside the corral. Anisotropic behavior can be found when comparing with the data taken for the same QD along the [100] direction (see Fig. 4(d) in the main text). (c) Particle-in-a-box simulation of the QD in (a), again finding reasonable comparability with the experiment above $E_F$, but stronger deviations below. (d) STM image of an artificial QD with $R = 6.15$ nm ($V_S = 5$ mV, $I_t = 100$ pA). (e) Set of normalized d$I$/d$V$ spectra measured along the line denoted in (d). Red (blue) refers to high (low) intensity. (f) Particle-in-a-box simulation of the QD in (d).